\documentclass[11pt,a4paper]{article}
\pdfoutput=1

\pdfoutput=1
\usepackage{jcappub}
\usepackage[utf8x]{inputenc}
\usepackage[title]{appendix}
\usepackage{graphicx}
\usepackage{color}
\usepackage{xfrac}
\usepackage{verbatim}
\usepackage{mathtools}
\usepackage{soul}

\title{Gravitationally produced Top Quarks and the Stability of the Electroweak Vacuum During Inflation}

\author[]{David Rodriguez Roman,}
\author[]{Malcolm Fairbairn}

\affiliation[]{Department of Physics, King's College London, Strand, London, WC2R 2LS, United Kingdom}

\emailAdd{david.rodriguez@kcl.ac.uk}             
\emailAdd{malcolm.fairbairn@kcl.ac.uk}                            

\abstract{In the standard model the (Brout-Englert-)Higgs quartic coupling becomes negative at high energies rendering our current electroweak vacuum metastable, but with an instability timescale much longer than the age of the Current Universe.  During cosmological Inflation, unless there is a non-minimal coupling to gravity, the Higgs field is pushed away from the origin of its potential due to quantum fluctuations. It is therefore a mystery how we have remained in our current vacuum if we went through such a period of Inflation. In this work we study the effect of top quarks created gravitationally during Inflation and their effect upon the Higgs potential using only General Relativity with minimal couplings and Standard Model particle physics. We show how the evolution of the Higgs field during Inflation is modified coming to the conclusion that this effect is non negligible for scales of Inflation close to or larger than the stability scale but small for scales where the Higgs is stable. Also, we briefly discuss the effect of other fermions to the Higgs instability.}

\begin{document}
\begin{flushleft}
	\hfill		  KCL-PH-TH/2018-33\\
\end{flushleft}
\maketitle

\section{Introduction}
\begin{figure}[h]
\begin{center}
\includegraphics[width=0.65\textwidth]{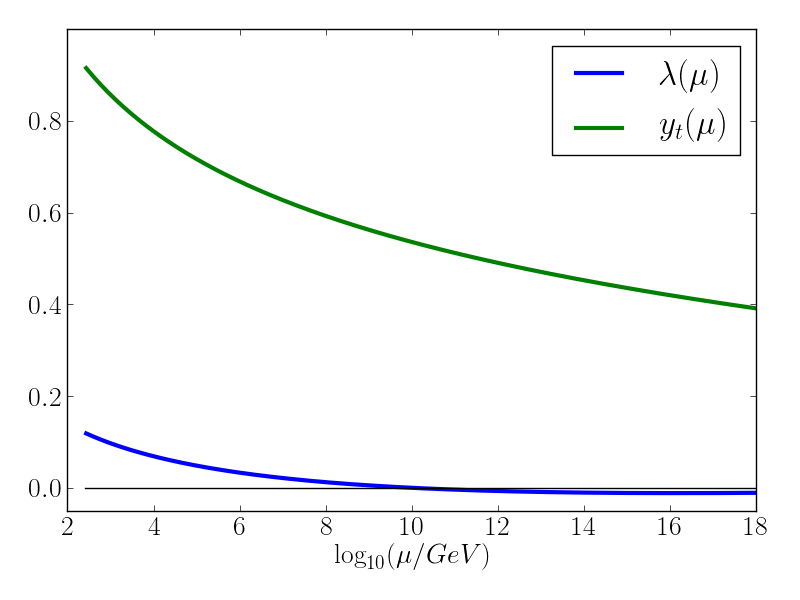}
\end{center}
\caption{\label{fig:running} \it Running of $\lambda$ and $y_t$ for $m_t= 173$ GeV, $\alpha_s = 0.1181, m_h = 125.18$ GeV.}
\end{figure}

The measurement of the actual Higgs and the top quark masses at the LHC and other colliders \cite{Khachatryan:2015hba,Aaboud:2016igd,Aad:2015zhl} leads to an interesting effect when one calculates their Renormalisation Group running in that the quartic Higgs self interaction coupling $\lambda$ becomes negative above around $10^{10}$ GeV \cite{Espinosa:2015qea,Buttazzo:2013uya,Degrassi:2012ry,EliasMiro:2011aa}.  This high energy scale cannot be probed at current colliders but is much smaller than the Planck mass and is in a region where all the couplings remain perturbative, so there is no reason not to take this extrapolation seriously. Taking the central observed values for the Higgs mass ($m_h$), the top quark mass ($m_t$), and the strong coupling constant ($\alpha_s$) from \cite{particledata}, a calculation \cite{Degrassi:2012ry} of the running of $\lambda$ and $y_t$ is shown in Figure~\ref{fig:running}

The implication of this is clear: in the absence of physics beyond the Standard Model affecting the running of the coupling constants, our current electroweak vacuum favours a metastable solution over an absolute stable vacuum \cite{Ellis:2009tp,Isidori:2001bm,Bezrukov:2012sa,Alekhin:2012py,Espinosa:2007qp,Kobakhidze:2013tn}.  Fortunately when one calculates the lifetime for tunneling into the true vacuum above $10^{10}$ GeV, one typically obtains numbers which are many orders of magnitude larger than the age of the Universe~\cite{Buttazzo:2013uya}, although it is still a subject of active research where new physics could modify the lifetime \cite{Branchina:2014usa,Branchina:2013jra,Branchina:2014rva,Branchina:2016bws,Branchina:2015nda,Bentivegna:2017qry}.  One might expect therefore that this unusual behaviour of the running at high scales is little more than a curiosity; however this situation changes when one considers the early Universe.

For several decades the leading hypothesis for the earliest stages of the evolution of the Universe has contained a period of cosmological Inflation where the scale factor expanded exponentially, solving many cosmological problems and explaining the origins of astrophysical structure formation across many orders of magnitude in physical scale \cite{Linde:1981mu,Albrecht:1982wi,Starobinsky:1980te}. While Inflation has its own fine tuning problems (see attempts to address and recast some of these here \cite{Fairbairn:2017krt}), there are not many compelling alternatives to Inflation which have a simpler or even equally simple mathematical consistency.

Fluctuations in the Higgs field during Inflation lead to stochastic growth in its expectation value which could push it to the region of instability at around $10^{10}$ GeV \cite{Starobinsky:1994bd} \footnote{It is usual to set the renormalization scale  $\mu$ to the expectation value of the Higgs $h$ when one considers effective potentials where the effects of loops are included as logarithmic corrections \cite{PhysRevD.7.1888,Espinosa:1995se}.}.  The Universe would then seemingly be overwhelmed by an anti-de Sitter (AdS) region which would subsequently collapse, allowing no possibility of us being here today \cite{Espinosa:2015qea,Fairbairn:2014zia,Kearney:2015vba,East:2016anr,Espinosa:2007qp}.  Because of this, there appears to be tight constraints upon the absolute scale of the expansion rate $H$ during Inflation in order to evade instability.  This corresponds in a one-to-one fashion upon the magnitude of primordial gravitational waves which might be generated during Inflation \cite{Frewin:1993dq,Kamionkowski:1996zd,Baumann:2009ds}, which is parametrized by the tensor to scalar ratio $r_T$.

What we propose in this paper is to  take into account for the first time the gravitational particle production of fermions during Inflation, in particular the top quark which has the strongest interaction with the Higgs field. 
The energy density of fermions produced during Inflation grows proportional to their mass \cite{Chung:2011ck}, and since top quarks have a Yukawa coupling $y_t$ of order unity, their mass is given by the Higgs vacuum expectation value (vev) $m_t\sim y_t \cdot h$. The interaction term in the Lagrangian of the SM for the case of the Higgs and the top fermions is
\begin{equation}
{\cal L}_\text{interaction} = y_t \frac{h}{\sqrt{2}} \bar{\psi} \psi
\end{equation}

So as the Higgs field is pushed to higher values, the mass of the top quarks will increase and the production of fermions will also increase, meaning that the contribution from the fermions $\bar{\psi}\psi$ to the Higgs potential will also rise. We aim to show that there are situations where this contribution to the potential can change the probability of ending in a catastrophic collapse during Inflation.

The paper is organised as follows, Section \ref{sec:unstable} reviews the instability of the electroweak vacuum during Inflation.  Section~\ref{sec:massivefermion} describes the particle production of massive fermions in a de-Sitter background and their subsequent modification of the Higgs potential in the case of top quarks. In Section~\ref{sec:results} we study the stability of the Higgs taking in consideration this effect before discussing the results in Section~\ref{sec:conclusions}

\section{The Instability of the Electroweak Vacuum during Inflation\label{sec:unstable}}

In this section we will review the normal arguments which explain why a period of Inflation is dangerous for the stability of the electroweak vacuum given the fact that the quartic coupling runs to negative values at high scales.

There is some discussion in the literature about the best choice of the scale $\mu$ and its relationship with the Higgs field expectation value $h$ when working with the Higgs in the Early universe. It was recently proposed \cite{Herranen:2014cua,Markkanen:2018bfx,Kearney:2015vba,East:2016anr} that when studying a quantum field in a curved space-time background, in order to cancel the logarithmic divergences that arise in the potential at one-loop order, the choice of the scale $\mu$ is different from the choice that is usually assumed for the same situation in a flat space-time background where $\mu \approx h$ is chosen \citep{Espinosa:2015qea}. In this work the results do not depend strongly on these two different choices of the scale but for definitiveness we choose to set the scale of the running as
\begin{equation}\label{eq:scale}
\mu=\sqrt{h^2+H^2},
\end{equation}
where $h$ is the Higgs vev and $H=\dot{a}/a$ is the Hubble parameter, although we will include an extension to our calculation to showcase the differences with the choice of scale $\mu=h$.\\

What is more widely agreed on is that during Inflation, short wavelength fluctuations behave as classical noise acting on the dynamics of the Higgs field on super-Hubble scales and these fluctuations can be described using the Langevin equation \cite{Starobinsky:1994bd,Hardwick:2017fjo}
\begin{equation}\label{eq:langevin}
\frac{d h}{d N_e}=-\frac{V'(h)}{3H^2} + \frac{H}{2 \pi} \xi
\end{equation}

Using this equation we can study how the expectation value of the Higgs field $\langle h^2\rangle$ evolves with $N_e$ - the number of e-folds of Inflation ($d N_e= d\ln a$, where $a$ is the scale factor). The evolution is due to a combination of two effects: the first is given by the classical equation of motion, where $V'(h)$ is the differentiation of the Higgs potential with respect to the Higgs vev, and the second is due to the stochastic noise, where $\xi$ is a Gaussian white noise with zero mean and unit variance. The Langevin equation is only valid for a light field $V''\ll H^2$.
If the Higgs is initially at the origin ($h=0$), the stochastic term dominates over the classical term and on average the Higgs vev after $N_e$ e-folds of Inflation would be 
\begin{equation}\label{eq:vev60}
\langle h^2 \rangle=\Big(\frac{H}{2\pi} \Big) ^2 N_e
\end{equation}

until the classical term becomes as large as the stochastic term, which in the classical picture occurs after $N_e=1/\sqrt{\lambda}$, and the Higgs would then acquire an equilibrium value given by
\begin{equation}\label{eq:vevstatic}
\langle h^2 \rangle = 0.13 \frac{H^2}{\sqrt{\lambda}}.
\end{equation}
This is valid only if $\lambda > 0$ (and constant). In the case that $\lambda$ is not positive, then the Higgs vev motion would be unbounded.  Note we are assuming here and throughout that the Higgs field starts at the origin 60 e-folds before the end of Inflation.  This assumption is somewhat important, but as long as $h$ starts somewhere below $H$ we expect very similar results.  If $h$ starts with a very high value, then a different kind of analysis would have to be performed. 

Therefore, even if we only assume $60$ e-folds of de-Sitter expansion, on average the value of the Higgs vev is going to be close to the energy scale of Inflation ($h \approx H$) and the running of the Higgs self interaction $\lambda ( \mu ) \approx \lambda ( H ) $. If the energy scale of Inflation is high enough, then the Higgs field would move into the unstable region; in particular, for 60 e-folds, the scale of Inflation should be about one order of magnitude smaller than the scale at the maximum of the potential  \cite{Espinosa:2015qea,Fairbairn:2014zia,Espinosa:2007qp}.\\
From the non detection of CMB polarisation associated with primordial gravitational waves ($r_T<0.12$) \cite{Ade:2015tva} we can set an upper bound on the energy scale of Inflation $H<10^{13}$ GeV and since the instability scale is around $\mu=10^{10}$ GeV \cite{Espinosa:2015qea}, we will focus on this energy interval $H=10^{9}-10^{13}$ GeV.\\

There are many possible alternative solutions to this problem of combining Inflation with the standard model.  However, unlike what we are proposing here, they all invoke new physics, the most obvious and well studied are a simple coupling between the Higgs field and the inflaton \cite{Fairbairn:2014zia,Ema:2017ckf,Gross:2015bea,Lebedev:2012zw} and a non-minimal coupling between the Higgs field and the Ricci Curvature \cite{Herranen:2014cua, Markkanen:2018bfx,Kohri:2016qqv,Kohri:2017iyl} or both at the same time \cite{Kohri:2016wof}. See also \cite{Goswami:2014hoa} for the effect of the Gibbons-Hawking radiation during Inflation on this problem or around an evaporating Black Hole in \cite{Kohri:2017ybt}.\\

Having explained the problem and shown that for Inflation with $H >10^{9}\text{GeV}$ the electroweak vacuum can be unstable, we now move on to consider the gravitational production of fermions and how they might change this situation.

\section{Massive fermion production}\label{sec:massivefermion}
In this section we consider how fermions, in our case top quarks, can be produced gravitationally and what effect they will have upon the Higgs potential. There are two general properties for the production of fermions that can be deduced independently of the details of the problem: First, fermions are conformally invariant, meaning that in the massless limit there is a conformal transformation from any Friedmann-Robertson-Walker (FRW) metric to Minkowski and therefore no particles are produced. Second, particle creation is exponentially suppressed for the case of {\it heavy} fermions ($m\gg H$) and large momenta. Both of these properties will be shown throughout the section. Here we follow closely the work of \cite{Chung:2011ck}.

It is now widely agreed that the fact that the definition of vacuum for a field in a curved space-time background is not unique leads to the production of particles \cite{Hawking:1974sw,Parker:1969au,Parker:1971pt}. In particular here we study a fermionic field that has been expanded in a helicity basis, $\psi=\sum_i a_i U_i+b_i^+V_i$, with $i=k,r$ and 
\begin{eqnarray}
U_{\vec{k},r}(\eta,\vec{x})=\frac{e^{i\vec{k}\cdot\vec{x}}}{(2\pi a)^{3/2}}
 \left( \begin{array}{cc} 
 u_A(k,\eta) h_{\hat{k},r}\\
  r u_B(k,\eta) h_{\hat{k},r}
   \end{array} \right)
\end{eqnarray}

where $\vec{k}$ is the momentum, $r= \pm 1$ is the helicity, $h_{\hat{k},r}$ is the helicity 2-spinor, and $u_A, u_B$ are the temporal parts of the field as a function of the conformal time $a d\eta = dt$ that solves the Dirac equation \cite{Chung:2011ck},
\begin{eqnarray}\label{eq:eom}
i\partial_\eta 
\left( \begin{array}{cc} 
 u_A(k,\eta)\\
  u_B(k,\eta)
   \end{array} \right)
=
\left( \begin{array}{cc} 
 a(\eta)m & k\\
  k & -a(\eta)m
   \end{array} \right)
\left( \begin{array}{cc} 
 u_A(k,\eta)\\
  u_B(k,\eta)
   \end{array} \right)
\end{eqnarray}

Since the choice of the orthonormal basis is not unique, we could define a different basis $\lbrace\tilde{U_i},\tilde{V_i}\rbrace$, where $\psi=\sum_i a_i U_i+b_i^+V_i=\sum_i \tilde{a_i} \tilde{U_i}+\tilde{b_i^+}\tilde{V_i}$.

The vacuum state is defined by $a_i|vac\rangle=b_i|vac\rangle=0$, so in the tilde basis, the number of particles measured over the initial (no-tilde) vacuum state is
\begin{equation}
\langle vac|\tilde{a_i^+}\tilde{a_i}|vac\rangle=\sum_j|\beta_{ij}|^2
\end{equation}

where the relation between the two vacuum states is linear and parametrised by the Bogoliubov coefficients ($\alpha_{\vec{k}}$ ,$ \beta_{\vec{k}}$): $\tilde{U}_{\vec{k}} = \alpha_{\vec{k}} U_{\vec{k}} + \beta_{\vec{k}} V_{\vec{-k}}$.

Defining the initial basis with the index `in' and the tilde basis in which we measure the number of particles of the initial vacuum state as `out', one obtains the following relation:
\begin{equation}
|\beta_k|=|u_A^{out}(k,\eta) u_B^{in}(k,\eta) - u_B^{out}(k,\eta) u_A^{in}(k,\eta)  |
\end{equation}

The `out' is set to be the instantaneous vacuum state (zeroth-order in adiabatic expansion), and it can be obtained by using the Wentzel-Kramers-Brillouin (WKB) approximation.
\begin{eqnarray}\label{eq:WKB}
\left( \begin{array}{cc} 
 u_A(k,\eta)\\
  u_B(k,\eta)
   \end{array} \right)^{WKB}
   = \alpha_k
   \left( \begin{array}{cc} 
 \sqrt{\frac{w+a m}{2w}}\\
 \sqrt{\frac{w-a m}{2w}}
   \end{array} \right) e^{-i \int^\eta w(\eta) d\eta}+
   \beta_k
   \left( \begin{array}{cc} 
 \sqrt{\frac{w-a m}{2w}}\\
 -\sqrt{\frac{w+a m}{2w}}
   \end{array} \right) e^{i \int^\eta w(\eta) d\eta}
\end{eqnarray}

where $w^2=k^2+m^2a^2$ and due to the normalization of the modes, $|\alpha_k|^2+|\beta_k|^2=1$. A vacuum state is defined as $\alpha=1$ and $\beta=0$.\\ This is a solution to (\ref{eq:eom}) in a Minkowski space-time where the scale factor is constant and there is no particle production, which is why it is called the instantaneous vacuum, because as the scale factor changes with time, this vacuum would measure a different number of particles.

The `in' state is the Bunch-Davies vacuum state for a perfect de-Sitter background solution to (\ref{eq:eom}), with $a(\eta)=-1/H\eta$~\cite{1978RSPSA.360..117B},\\
\begin{eqnarray}\label{eq:vacuumInflation}
\left( \begin{array}{cc} 
 u_A(k,\eta)\\
  u_B(k,\eta)
   \end{array} \right)^{in}
   =\sqrt{\frac{\pi}{4}k\eta} 
   \left( \begin{array}{cc} 
e^{+\frac{\pi m}{2 H}} H_{\frac{1}{2}-i\frac{m}{H}}^{(1)} (-k \eta) \\
e^{-\frac{\pi m}{2 H}} H_{\frac{1}{2}+i\frac{m}{H}}^{(1)} (-k \eta) 
   \end{array} \right)
\end{eqnarray}

In the limit $a \rightarrow 0$ ($\eta \rightarrow - \infty$) agrees with the WKB solution (\ref{eq:WKB}) at that time, therefore at the beginning there are no particles since both states coincide with $\alpha=1$ and $\beta=0$.\\
In order to not create extra particles from the sudden measurement of particles in the instantaneous vacuum (\ref{eq:WKB}), we introduce a smooth exit from Inflation into Minkowski space-time such that $H_\eta(\eta)=H (1-\tanh ((\eta-\eta_i)/\eta_0))/2$, where $\eta_i$ is the time at which Inflation ends, $H$ is the value of the Hubble parameter during Inflation and $\eta_0$ is the speed of the transition. Then (\ref{eq:vacuumInflation}) is the solution to (\ref{eq:eom}) for $\eta \ll \eta_i$ and (\ref{eq:WKB}) is the solution at $\eta \gg \eta_i$ where we can unequivocally define the number of particles created during the de-Sitter period of expansion of the universe.\\
The speed of the transition is set to $\eta_0=1/H$, the natural scale for Inflation. In the low limit mass, $m/H \ll 1$, the calculation is unaffected by the speed of the transition. But for masses $m/H \geq 1$ if the transition is faster, $\eta_0 \ll 1/H$, then more particles would be created because of the sudden change in the scale factor, and if $\eta_0 \gg 1/H$, the transition happens too slow and heavy fermions would be diluted leading to a smaller number of particles being produced. For a more exhaustive study of the effect of the speed of the transition we refer the reader to the work done for the case of scalar fields in \cite{Chung:1998zb}.\\

\begin{figure}[h]
\begin{center}
\includegraphics[width=0.65\textwidth]{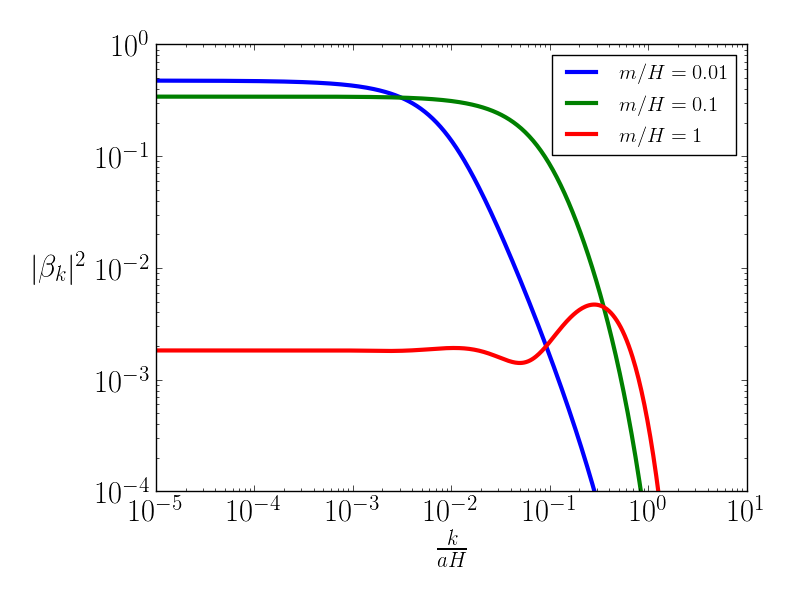}
\end{center}
\caption{\label{fig:betas} \it Plot of $| \beta_k |^2$ as a function of $k/aH$ for different masses. If the fermions are light, the spectrum can be approximated as $1/2$ up to $k/a=m$; for the heavy fermions the spectrum is suppressed as $1/(1+e^{2\pi m/H})$.}
\end{figure}

The production of {\it heavy} fermions, $m \gtrsim H$, is exponentially suppressed by their mass $(1+e^{2\pi m/H})^{-1}$ but for the case of {\it light} fermions, $| \beta_k |^2=1/2$ is constant up to $k/a=m$ as shown in Figure~\ref{fig:betas}.

The quantity we are interested in is the expectation value of an initial vacuum state for the product $\langle \bar{\psi} \psi \rangle$, and using (\ref{eq:WKB}) this takes the form
\begin{equation}
\langle \bar{\psi} \psi \rangle = \int \frac{d^3 k_p}{2\pi^3}\frac{m}{w_p}|\beta_k|^2
\end{equation}
where the subscript $p$ stands for  $\it{physical}$ quantities, so $k_p=k/a,w_p=w/a$. Also the piece in the product coming from the initial vacuum and an oscillatory term has been discarded, as it has been done as well in \cite{Peloso:2000hy, Chung:2013rda}; in the literature this is called normal ordering or renormalization of the product.

\begin{figure}[h]
\begin{center}
\includegraphics[width=0.65\textwidth]{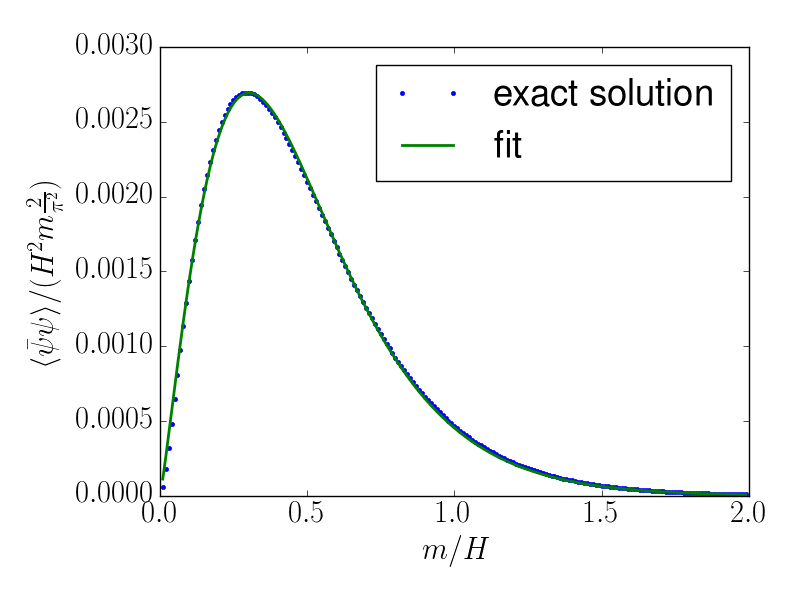}
\end{center}
\caption{\label{fig:vevs} \it Plot of $\frac{\langle \bar{\psi} \psi \rangle \pi^2}{2 H^2 m}$ as a function of $m/H$. If the fermions are light, $\langle \bar{\psi} \psi \rangle \propto m^{2.2}$ up to $m/H=0.49$, above which it is exponentially suppressed.}
\end{figure}

In this way we can obtain the expectation value for a massive fermion during Inflation as a function of its mass as shown in Figure~\ref{fig:vevs}

\begin{equation}\label{eq:vev}
\langle \bar{\psi} \psi \rangle = H^3 \frac{m}{H} \frac{2}{\pi^2} \frac{0.063 \big(\frac{m}{H}\big)^{1.22}}{e^{4.92 \frac{m}{H}}+1} 
\end{equation}

\subsection{Addition to the Higgs potential}
\begin{figure}[h]
\begin{center}
\includegraphics[width=0.65\textwidth]{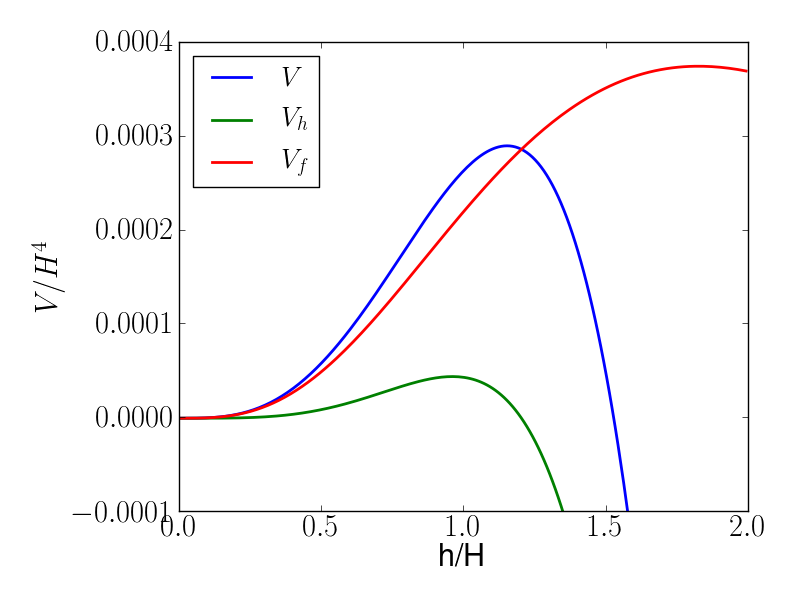}
\end{center}
\caption{\label{fig:plotmu} \it Plot of $V/H^4$ as a function of $h/H$ for $H=10^{10} \text{GeV}$. With the additional effect of the fermionic contribution to the potential $V_f$, the total potential $V=V_h+V_f$ has a barrier which is five times higher.}
\end{figure}

The full Lagrangian that determines the dynamics of the Higgs field is
\begin{equation}
{\cal L}_\text{Higgs \& top} =\frac{1}{2} \partial_\mu h \partial ^\mu h - \frac{\lambda}{4} h^4 -3 y_t \frac{h}{\sqrt{2}} \bar{\psi} \psi+ 3 i \bar{\psi} \gamma^{a} e_{a}^{\mu} \nabla_{\mu}\psi
\end{equation}
where the spin-$\frac{1}{2}$ covariant derivative with vierbein dependent spin-connection, $\omega_{\mu}^{\alpha \beta}$, is defined as $\nabla_{\mu}\psi=\partial_{\mu}\psi+\frac{1}{8}\omega_{\mu}^{\alpha \beta}[\gamma_\alpha,\gamma_\beta]\psi$, $e_{a}^{\mu}$ is the vierbein, and $\gamma_\alpha$ are the standard Minkowski space-time Dirac matrices. The mass of the fermions is of course explicitly given by the Higgs expectation value.  This coupling through the Yukawa coupling $y_t$ also leads to a term in the equation of motion for $h$ which is proportional to $\bar{\psi} \psi$; and therefore the fermions change its dynamics, and the factor $3$ comes from the color charge of the quarks in the Standard Model (as pointed out in \cite{Franciolini:2018ebs}).
The addition to the Higgs potential coming from the production of fermions is (using the result obtained in (\ref{eq:vev}))
\begin{eqnarray}\label{eq:potentialHiggsfull}
V(h)=V_\text{h}+V_\text{f}=\frac{\lambda}{4} h^4 + 3 y_t \frac{h}{\sqrt{2}} \bar{\psi} \psi = \frac{\lambda}{4} h^4 +3 H^4 \frac{0.013 (y_t \frac{|h|}{\sqrt{2}H})^{3.22}}{e^{4.92 (y_t \frac{|h|}{\sqrt{2}H})}+1}
\end{eqnarray}
The condensate created from the production of fermions changes the Higgs potential, adding an extra term that peaks at $h_{\text peak} = 0.96 H/y_t$. At that value the potential has the value
\begin{equation}
\frac{V (h_{\text peak})}{H^4}=\frac{\lambda}{4}\frac{0.84}{y_t^{4}} +0.00037
\end{equation}
So the contribution from the fermions to the Higgs potential can dominate if 
\begin{equation}\label{eq:ytl}
y_t  > 4.8 \cdot \lambda^{1/4}
\end{equation}
As can be seen in Fig.\ref{fig:plotmu}, the height of the barrier is increased and there is a visible shift in the scale of the instability; however, later we will see that this has a disappointingly small effect upon the overall probability of becoming unstable. Note that the effect of the fermion back reaction dominates the potential when the criterion (\ref{eq:ytl}) is fulfilled, i.e. very close to the point where $\lambda \sim 0$, at that scale $y_t \sim 0.5$, so $V_\text{f}$ peaks at $h/H \sim 2$.\\
If we were to study another fermion with a different Yukawa coupling to the SM Higgs field, assuming that Inflation occurs at a low energy scale where $\lambda \sim 0.1$, we would need a Yukawa coupling bigger than $y>2.7$ in order for the fermions to dominate (after 60 e-folds of fermion condensate production). So the bigger the Yukawa coupling the bigger the effect, which is why we have been focusing on the top quarks throughout this paper.  We note, however, that a fermion with a larger Yukawa coupling would destabilise the vacuum at a much lower value of the Higgs field.

The main difference in comparison with the calculation in Section~\ref{sec:massivefermion} is that the fermion mass is not a constant but now depends on the Higgs vev. The relevant term in (\ref{eq:eom}) is $m a(\eta)= y_t \frac{h}{\sqrt{2}} a(\eta)$ which clearly varies as $h$ changes. We need to establish if assuming that the mass is constant is a good approximation so as to trust our calculation.  To do this we need to compare the variation with time of the Higgs field with that of the scale factor and ensure that $\frac{h'}{h} \ll \frac{a'}{a}$, where $' \equiv \frac{d}{d \eta}$.  We look at this assumption in more detail in Appendix~\ref{ap:massconstant}. The variation with time of the Yukawa coupling is not considered since it would come from its running, but it should be close to zero since we are assuming close to perfect de-Sitter and $\mu \approx H $. If we assume that the renormalisation scale $\mu$ is given by $h$ and not $H$, then the only difference will be a larger value of $y_t$ during the first e-fold of Inflation, but during that time the top quarks are almost massless ($h \ll H$) and then their production negligible, so we do not expect the calculation to be sensitive to this choice. \\
The Higgs will jump stochastically due to quantum fluctuations, and in one e-fold the size of a single quantum jump is $H/2\pi$ \cite{Guth:2007ng}; therefore,
\begin{equation}
\frac{d h}{d N_e} = \frac{H}{2\pi}
\end{equation}
and from the definition of the Hubble parameter $a'=H a^2$.\\

Then the assumption of having a constant mass in this case is rewritten such that
\begin{equation}\label{eq:massconstant}
\frac{h'}{h} \ll \frac{a'}{a} \Rightarrow \frac{H}{2\pi}  \ll h ,
\end{equation} 
and from (\ref{eq:vev60}) this is true after the first e-fold. Before that, the Higgs vev is close to zero, making the fermion almost massless, and since the production of the fermions is proportional to their mass, it is safe to neglect the production from the time when the (\ref{eq:massconstant}) does not hold.

We have shown in this section how the gravitationally produced top quarks will contribute to the Higgs potential. In the next section we will study how this might affect the stability of the electroweak vacuum during Inflation.

\section{Stability study}\label{sec:results}
The Higgs field during Inflation is moving stochastically. Even though the variance of the field is given by (\ref{eq:vev60}) or (\ref{eq:vevstatic}), the probability distribution function extends to infinity. Therefore there is a possibility of going over the barrier and ending up in an anti-de Sitter region. Since this is not the case in our current Hubble horizon, which is composed of $e^{3 N_e}$ causally independent regions, we need to impose the condition that the probability of going over the barrier is, at least, smaller than $e^{3 N_e}$ because none of these regions can be in an anti-de Sitter space-time.\\
We do not study the evolution of the Higgs field after Inflation and the possibility that even if the Higgs goes over the barrier, thermal effects can make it go back to the False vacuum. For more details concerning this, the reader should consult \cite{Espinosa:2015qea}.\\

\begin{figure}[h]
\minipage{0.48\textwidth}
\includegraphics[width=\textwidth]{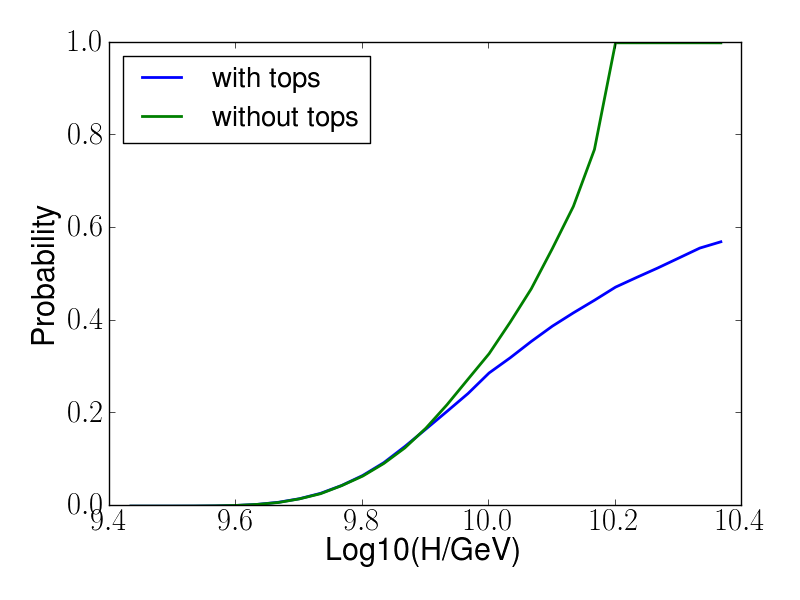}
\caption{Probability of going over the barrier after 60 e-folds for a renomalisation scale $\mu^2=h^2+H^2$.}\label{fig:probamuhh}
\endminipage\hfill
\minipage{0.48\textwidth}
\includegraphics[width=\textwidth]{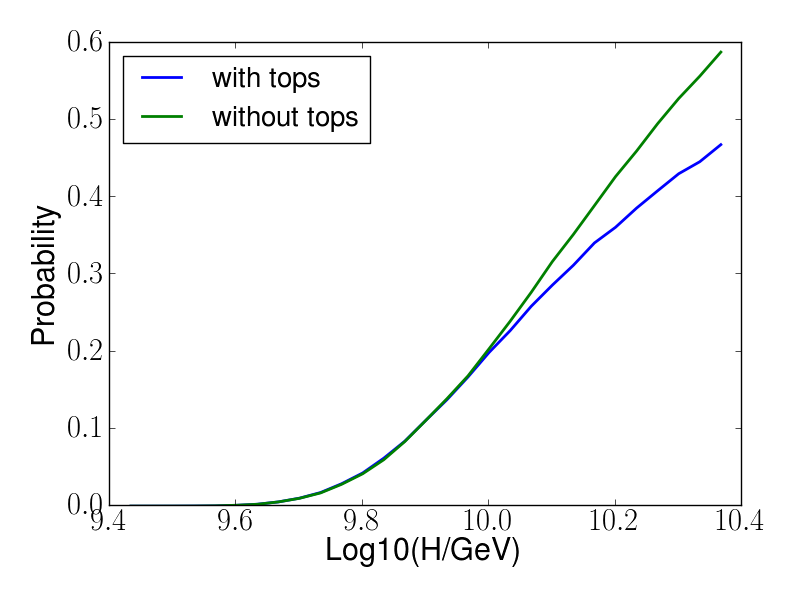}
\caption{Probability of going over the barrier after 60 e-folds for a renomalisation scale $\mu=h$.}\label{fig:probamuh}
\endminipage
\end{figure}

First, we solved numerically the Langevin equation (\ref{eq:langevin}) using the modified potential (\ref{eq:potentialHiggsfull}) and obtained the probability that after 60 e-folds of Inflation, the Higgs field would have gone over the barrier, using both prescriptions to determine the scale, $\mu^2=h^2+H^2$ (Fig. \ref{fig:probamuhh}) and $\mu=h$ (Fig.\ref{fig:probamuh}). To get reliable statistics we simulated $10^5$ realizations. The way we determined if the Higgs goes over the barrier is, after 60 e-folds, if $V'(h_{60})<0$, it has gone over the barrier and in the opposite case, it has not.

For the choice of scale $\mu^2=h^2+H^2$ and $H>10^{10.2}$GeV, the Higgs is always unstable (probability is always one). This is because the value of $\lambda$ is always negative and the Higgs, independently of its vev, ends up in an AdS vacuum. Once the production of top quarks is taken into account, the probability is smaller than one since even though the value of $\lambda$ is negative, there is still a barrier generated by the top quarks preventing the Higgs from ending in an AdS region.\\
Comparing both plots it can be seen that for the prescription $\mu^2=h^2+H^2$, one is more likely to end in an AdS region for values of the Hubble parameter close to $10^{10}\text{GeV}$ than in the case where the scale is just given by the Higgs vev. It makes sense since in the former case there is a minimum value for the scale $\mu=H$, and therefore the value of $\lambda$ is smaller and closer to zero, independently of the Higgs vev. In this case, the Higgs feels "less" the potential and can acquire a larger vev during 60 e-folds of Inflation.\\
Also it can be seen that if the scale of Inflation is reduced, there is almost no difference adding the top quarks to the potential - since the contribution from the fermions is determined by the scale of Inflation, the smaller the scale, the smaller the effect. But for the cases where it is important, the top quarks can reduce that probability by up to 50$\%$ in the prescription $\mu^2=h^2+H^2$ and 10$\%$ for $\mu=h$.\\

The stability condition is that the probability $P<e^{-3 N_e}$. For 60 e-folds it is not possible to study it numerically since the number of realisations that we would need is of the order $e^{3 N_e}$. Instead we estimate the effect from the fermions analytically.\\

As can be inferred from Fig.\ref{fig:probamuhh} and \ref{fig:probamuh}, the effect from the top quarks for values of the scale of Inflation $H < 10^{10}$GeV is very small, so we can treat this effect perturbatively.\\
Following the work in \cite{Espinosa:2007qp,Markkanen:2018bfx} for the study of the Higgs instability without the top quarks, it was shown that for values of $H<h_{max}$ ($h_{max}$ is the Higgs vev at the maximum of the potential), within 60 e-folds of Inflation, the Higgs acquires a constant variance, eq. (\ref{eq:vevstatic}). Since the time it takes to reach an equilibrium distribution is given by $N_e=1/\sqrt{\lambda}$ and we are studying the situation where $H<h_{max}$, $\lambda$ is never so small that the time it would take to reach the equilibrium distribution was longer than 60 e-folds. In this situation, the stochastic motion is compensated with the gradient of the Higgs potential and acquires an equilibrium distribution. Once an equilibrium has been reached $P(h>h_{max})=e^{-\frac{8 \pi^2}{3 H^4}V_\text{max}}$, so the stability condition is
\begin{eqnarray}
\frac{8 \pi^2}{3 H^4}V_\text{max}> 3 N_e
\end{eqnarray}
where $V_\text{max}$ is the value of the Higgs potential at its maximum.\\

\begin{figure}[h]
\begin{center}
\includegraphics[width=0.65\textwidth]{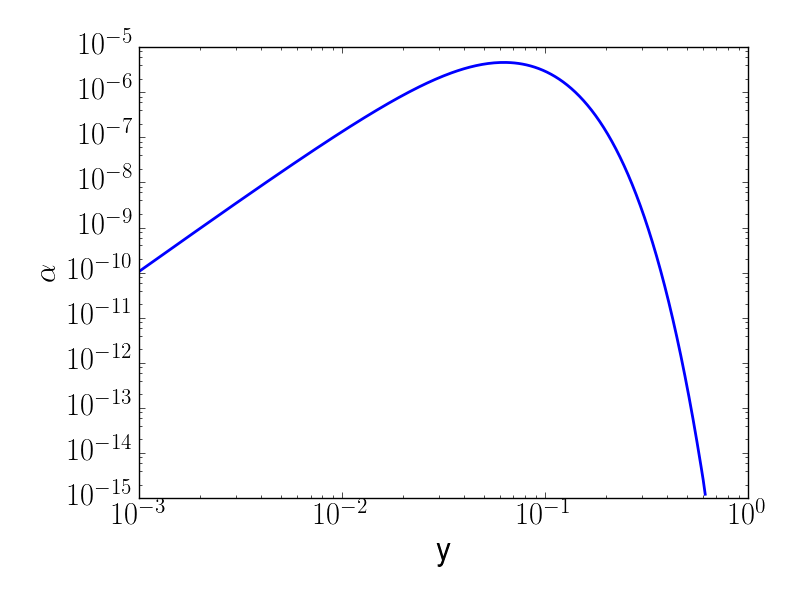}
\end{center}
\caption{\label{fig:ploty} \it Plot of $\alpha$ as a function of $y(\mu=h_\text{max})$.It peaks at $y=0.062$.}
\end{figure}

The difference from previous studies is that we now consider what is the effect of the top quarks, using (\ref{eq:potentialHiggsfull}) evaluated at a scale $\mu=h_\text{max}$ for both prescriptions, since we are studying the cases where $H<h_\text{max}$. The value of the Higgs self interaction $\lambda$ close to $h_\text{max}$ is estimated as in \cite{Espinosa:2015qea}, $\lambda=0.08/(4 \pi)^2$. It is helpful to define $x=h_\text{max}/H$, which for the case of the Higgs without the addition of the top quarks to the potential is $x_\text{h}=15.24$ for 60 e-folds of Inflation. Giving us a bound on the scale of Inflation to make the Electroweak vacuum stable, $H<8 \cdot 10^{8}\text{GeV}$. The effect of the fermions coupled to the Higgs is parametrized like
\begin{equation}\label{eq:xalpha}
\frac{x_\text{h+f}}{x_\text{h}}=1-\alpha
\end{equation}
where there is a minus sign since this effect makes the Higgs more stable.
In general this can be generalized to any quark with a Yukawa coupling to the Higgs. The difference in the stability scale induced by this effect is shown in Fig.(\ref{fig:ploty}). The value of the top Yukawa coupling at the instability scale is $0.5$ from Fig.(\ref{fig:running}), and therefore $\alpha_\text{top} \sim 10^{-13}$.

The maximum difference on the stability scale comes from a Yukawa coupling of $y \sim 10^{-2}$, i.e. bottom quarks, and that would be $\alpha_\text{bottom} \sim 10^{-6}$, still a small difference but orders of magnitude larger that the effect from the top quarks.\\
The effect of the bottom quarks is larger than the top because the scale of Inflation is smaller than $y(\mu=h_\text{max}) h_\text{max}$ since the effects of the fermions in the potential peaks at $h_{\text peak}$, but in situations where $H  > y(\mu=h_\text{max}) h_\text{max}$ the larger the Yukawa coupling, the larger the effect up to $h_{\text peak}=h_\text{max}$. Overall if there were to be a significant change to the study of the Electroweak vacuum it would come from the top quarks since, proportionally, they modify the Higgs potential the most (see Fig.(\ref{fig:probamuhh}),(\ref{fig:probamuh})). Although in a scenario where the scale of Inflation is small enough such that there is not a problem with the stability of the Higgs, then the biggest effect would come from the bottom quarks despite being a tiny effect.

\section{Discussion}\label{sec:conclusions}
With only Standard model particle physics, the Higgs field $h$ seems to become unstable at renormalisation scale $\mu>10^{10} GeV$ and from the non detection of primordial tensor perturbations we know that during Inflation $H<10^{13} GeV$.  If Inflation occurs with a value of $H$ within this range, there is generically a problem with the stability of the Higgs field.

In this work we have shown how without the addition of physics Beyond the Standard Model the gravitational production of quarks during Inflation changes the Higgs potential in such a way as to make it more stable.

Since the Higgs vev gives the quarks their mass, if it obtains a large value during Inflation, the fermions become relevant to the Higgs potential as shown in (\ref{eq:potentialHiggsfull}). This contribution can be large enough to prevent the Higgs from being pushed into the true vacuum during Inflation in borderline cases (Figures \ref{fig:probamuhh} and \ref{fig:probamuh}).

It is also clear from the stability study (Sec.\ref{sec:results}) that since we have not added anything new to the SM and there are no free parameters, there is no apparent possibility of improving these results. At the very least it is possible to extend the stability of the Higgs a little bit (\ref{eq:xalpha}).

Nevertheless we find this an interesting and noteworthy effect. Possible future extensions of this work would be looking at the effect of fermions beyond the standard model to see if there is any way that they would change the situation.  In summary, in the Standard Model the Higgs field seems to be unstable during Inflation, but slightly less unstable than before this effect is taken into account.

\acknowledgments
It is a pleasure to thank Tommi Markkanen, Toni Riotto and Tommi Tenkanen for useful conversations. This project was funded by the European Research Council through the project DARKHORIZONS under the European Union's Horizon 2020 program (ERC Grant Agreement no.648680). The work of MF was also supported by the STFC.   
\begin{appendices}
\section{Justification of the constant mass approximation}\label{ap:massconstant}
The calculation for the production of fermions in Section~\ref{sec:massivefermion} was done under the assumption that the mass, $m$, of the fermions is constant; i.e. it does not change with time. Here we aim to justify that this was a reasonable assumption.  In the case of the top quark, the mass is given by the Higgs vev as
\begin{equation}
m= y_t \frac{h}{\sqrt{2}}
\end{equation}
A way of solving (\ref{eq:eom}) is to obtain a second order ODE for $u_A$ and/or $u_B$,
\begin{equation}
u''_B+u_B (k^2-i(m a(\eta))'+(ma(\eta))^2))=0.
\end{equation}
The frequency of this ODE is $\omega^2=k^2-i(m a(\eta))'+(ma(\eta))^2)=k^2-i(y_t \frac{h}{\sqrt{2}} a(\eta))'+(y_t \frac{h}{\sqrt{2}}a(\eta))^2)$.

We want to compare the frequency with a time dependent value of $h(\eta)$,
\begin{equation}
\omega_1^2(\eta)=k^2-i(y_t \frac{h(\eta)}{\sqrt{2}} a(\eta))'+(y_t \frac{h(\eta)}{\sqrt{2}}a(\eta))^2)
\end{equation}
with the expression where $h$ is a constant,
\begin{equation}
\omega_2^2(\eta)=k^2-i(y_t \frac{h}{\sqrt{2}} a(\eta))'+(y_t \frac{h}{\sqrt{2}}a(\eta))^2)
\end{equation}
The way we compare them is 
\begin{equation}
\frac{\omega_1^2(\eta)'}{\omega_2^2(\eta)'}=1+e
\end{equation}
where if the approximation of making the mass constant for the case of the top quark is good, then $e \ll 1$.\\

The variation of the Higgs vev with time is $h' = a(\eta) \frac{H^2}{2\pi}$.\\
After $N_e \approx 1.2$ e-folds, $e<1$ and it is a good approximation to consider the mass of the fermions constant. So we can conclude that $\frac{H}{2 \pi} \ll h$ let us assume that the mass is constant in comparison with the variation of the scale factor.
Analytically this can be seen as 
\begin{eqnarray}
\omega_1^2(\eta)=k^2+a^2H^2(\frac{y_t^2 h^2}{2H^2}-i\frac{y_t}{2\sqrt{2}\pi}-i\frac{y_t h}{\sqrt{2} H})\\
\omega_2^2(\eta)=k^2+a^2H^2(\frac{y_t^2 h^2}{2H^2}-i\frac{y_t h}{\sqrt{2} H})
\end{eqnarray}
Knowing that $y_t < 1$ if $\frac{H}{2 \pi} \ll h$, then $\omega_1 \approx \omega_2$.  We therefore argue that for the vast majority of the e-folds of Inflation, the approximation we have taken is a good one, and for the small period of Inflation where the mass is very small, the production of fermions is anyway negligible, as argued in the main body of the article.

\end{appendices}


\bibliography{fermions}

\end{document}